\documentclass{article}
\usepackage[utf8]{inputenc}
\usepackage{amsmath}
\usepackage{graphicx}
\usepackage{hyperref}
\usepackage{geometry}
\usepackage{listings}
\usepackage{float}
\usepackage[numbers]{natbib}
\usepackage{listings}
\usepackage{xcolor}
\title{Intent-Aware Authorization for Zero Trust CI/CD}
\author{Surya Teja Avirneni \\
Cloud Platform and Security Architect \\
IEEE Member, ISC2 Certified, ACM Member \\
United States}
\date{April 2025}

\begin{document}

\maketitle

\begin{abstract}
In Zero Trust CI/CD systems, identity is only the beginning. Once a job or workload is authenticated, a decision must still be made about what actions it is allowed to perform, when, and under what conditions. This paper explores how policy engines embedded within credential brokers can support intent-aware authorization—evaluating not just identity but also justification, timing, and workload context. We describe a control loop model where SPIFFE-issued identities are passed into policy evaluators such as OPA or Cedar, and decisions are made based on runtime context and declared intent. We present architectural patterns that support fine-grained access, human approvals, and token scoping. The paper builds on our earlier work on SPIFFE-based workload identity and credential brokers, and focuses here on the policy logic that enables least-privilege, auditable access. This work is the third in a series on Zero Trust CI/CD.
\end{abstract}

\section{Introduction}

Most CI/CD platforms today are still driven by a narrow view of identity—either a cloud role assumed via OIDC federation or a service account mounted into a runner. But access is not just about identity. A secure deployment pipeline must also understand \emph{intent}: what is the purpose of access? Has it been approved? Is it happening in the right context, at the right time?

In previous papers, we established SPIFFE-based identity as a secure alternative to secrets and static credentials \cite{avirneni2025spiffe}, and introduced credential brokers that enforce runtime access control based on identity and policy \cite{avirneni2025broker}. This paper extends that foundation by focusing on how policy engines can make access decisions that are not only identity-aware, but also intent-aware.

In real-world platforms, deployments often require approvals, SLA considerations, or compliance constraints. Traditional IAM roles and ABAC policies are poorly suited to these cases. They do not evaluate justification, and they rarely operate in real time. This creates an enforcement gap—one that cannot be solved by stronger identity alone.

This paper introduces the concept of \emph{intent-aware authorization}, where policies consume both identity and justification to decide access dynamically. We explain how policy engines like OPA and Cedar can model approvals, time-based constraints, and runtime metadata as part of a control loop. We also show how these decisions can be integrated into CI/CD credential brokers, completing the Zero Trust architecture.

\section{What is Intent-Aware Authorization?}

In traditional IAM systems, access control decisions are based on predefined role or group mappings. These models—RBAC (Role-Based Access Control) and ABAC (Attribute-Based Access Control)—assume that access privileges can be assigned ahead of time, regardless of when or why access is being requested. In Zero Trust CI/CD, this assumption no longer holds.

\textbf{Intent-aware authorization} adds a critical missing dimension: \textit{why} access is being requested. It treats access as a contextual decision that must account for human approvals, workload metadata, SLA conditions, deployment type, and business justification.

For example, two CI jobs might share the same SPIFFE identity, but only one may have a recorded approval for production deployment. Similarly, emergency patches may require elevated access with a time-bound override token issued by an incident manager.

\subsection*{Why Static IAM Falls Short}

Static IAM policies do not consider:
\begin{itemize}
  \item \textbf{Justification context}: Why is access needed? Is it tied to an approved change ticket?
  \item \textbf{Temporal constraints}: Is this deployment occurring during an allowed window?
  \item \textbf{SLA-driven gating}: Does the job relate to a failed availability check?
  \item \textbf{Human-in-the-loop approvals}: Is there an active approval from an SRE or release manager?
\end{itemize}

Intent-aware authorization brings these dimensions into the policy decision point.

\subsection*{CI/CD Example: Production Deployment Approval}

A real-world scenario:

\begin{itemize}
  \item CI job: \texttt{spiffe://ci/org/release}
  \item Action: Push container to production
  \item Context:
    \begin{itemize}
      \item Git commit approved and merged
      \item PagerDuty override token attached
      \item SLA breach recorded on affected service
    \end{itemize}
  \item Policy: Allow only if all three conditions are true
\end{itemize}

This style of policy evaluation cannot be encoded in IAM roles or predefined group-based access models. It requires runtime signals and intent awareness at the point of access.

\subsection*{Relation to ABAC: From Attributes to Intent}

ABAC systems traditionally decide access based on a set of attributes about the subject, resource, action, and environment. While this model is flexible, most real-world ABAC implementations use static tags—such as environment = “prod” or team = “billing”—to make decisions. This misses the dynamic nature of CI/CD workflows.

\textbf{Intent-aware authorization extends ABAC} by introducing runtime signals and justification tokens as first-class attributes. For example:
\begin{itemize}
  \item \texttt{input.justification.approval = "change\_12345"}
  \item \texttt{input.deployment.window = "offpeak"}
  \item \texttt{input.sla\_breach = true}
\end{itemize}

These attributes are ephemeral, externally validated, and tied to the specific execution context—not hardcoded into the identity or environment.

Rather than replacing ABAC, intent-aware systems build on its foundations, adding dynamic attributes, temporal logic, and external state evaluation. This evolution supports precise, policy-driven access in CI/CD pipelines where context and justification matter more than static roles or tags.

\section{Control Loops in Authorization}

Traditional access control models treat authorization as a one-time gate: a request arrives, a policy is evaluated, and access is either granted or denied. In Zero Trust CI/CD systems, this binary, stateless decision is insufficient.

\textbf{Intent-aware authorization} models access as a \textit{continuous control loop}, where each access decision is influenced by identity, justification, runtime context, and feedback signals. This loop supports revocation, dynamic approval, and real-time policy enforcement—particularly important in CI/CD environments where jobs are ephemeral and operational conditions change rapidly.

\vspace{1em}
\noindent\textbf{Authorization as a Runtime Control Loop:}
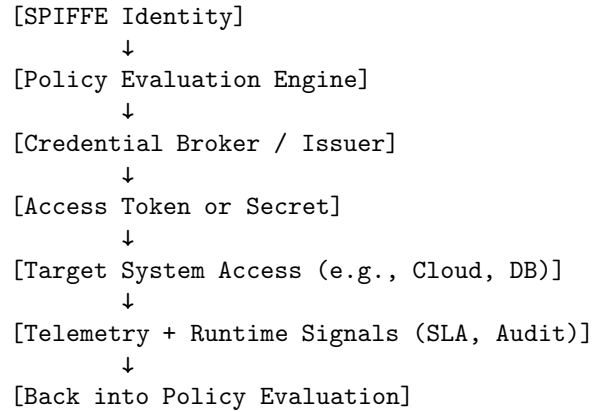
\begin{verbatim}
[SPIFFE Identity]
        ↓
[Policy Evaluation Engine]
        ↓
[Credential Broker / Issuer]
        ↓
[Access Token or Secret]
        ↓
[Target System Access (e.g., Cloud, DB)]
        ↓
[Telemetry + Runtime Signals (SLA, Audit)]
        ↓
[Back into Policy Evaluation]
\end{verbatim}
\vspace{1em}

This structure formalizes intent-aware enforcement as a feedback-driven system. Each access request is evaluated based on real-time conditions, and access may be revoked or denied in subsequent iterations if policy violations or environmental changes are detected.

\subsection*{Contrasting Static and Loop-Based Authorization}

In traditional IAM models (RBAC/ABAC), access is granted through predefined mappings:

\begin{verbatim}
[Job Identity] → [IAM Role] → [Static Access Grant]
\end{verbatim}

In intent-aware systems, access is granted only after dynamic, real-time evaluation:

\begin{verbatim}
[Job Identity + Context] → [Dynamic Policy Evaluation] → [Just-in-Time Access]
\end{verbatim}

This distinction is crucial in Zero Trust CI/CD: jobs are short-lived, policies must reflect current state, and context must evolve across pipeline stages.

\subsection*{Examples of Loop Triggers}

Intent-aware control loops respond to signals such as:
\begin{itemize}
  \item Expiration of an incident override token
  \item Approval withdrawn on a change ticket
  \item SLA status changing from "critical" to "stable"
  \item Time-of-day restrictions or maintenance windows
  \item Security alert on the source environment
\end{itemize}

Each of these can trigger re-evaluation, denial of further access, or revocation of previously granted credentials.

\subsection*{Rego (OPA) Example}

\begin{lstlisting}
package authz

default allow = false

allow {
  input.spiffe_id == "spiffe://ci/org/deploy-job"
  input.action == "push"
  input.resource == "s3://prod-release-artifacts"
  input.justification.status == "approved"
  within_maintenance_window(input.time)
}

within_maintenance_window(time) {
  time >= "02:00"
  time <= "05:00"
}
\end{lstlisting}

\textbf{Sample Input (runtime context):}

\begin{lstlisting}
{
  "spiffe_id": "spiffe://ci/org/deploy-job",
  "action": "push",
  "resource": "s3://prod-release-artifacts",
  "justification": { "status": "approved" },
  "time": "02:15"
}
\end{lstlisting}

\subsection*{Cedar Policy Example}

\begin{lstlisting}
permit(
  principal == workload::"spiffe://ci/org/deploy-job",
  action == action::"publish",
  resource == artifact::"prod-artifact"
)
when {
  context.justification == "approved_change_ticket" &&
  context.override == false &&
  context.timestamp >= "2025-04-20T02:00:00Z" &&
  context.timestamp <= "2025-04-20T05:00:00Z"
};
\end{lstlisting}

\subsection*{Python Broker Simulation (Control Loop)}

\begin{lstlisting}
def evaluate_access(input_data):
    if input_data["spiffe_id"] != "spiffe://ci/org/deploy-job":
        return False
    if input_data["justification"]["status"] != "approved":
        return False
    return is_within_window(input_data["time"])
\end{lstlisting}

This code simulates runtime re-evaluation on every access request. If conditions change, access is denied—even for a previously valid SPIFFE ID.

\subsection*{Feedback-Driven Authorization}

By treating policy as a loop—not a static rule set—CI/CD systems gain:
\begin{itemize}
  \item \textbf{Fine-grained revocation} of active credentials
  \item \textbf{Continuous enforcement} based on live environment state
  \item \textbf{Observability-aware gating}, tying access to monitoring or audit conditions
\end{itemize}

These loops enforce Zero Trust principles: \textit{never trust, always verify}—and reverify with every credential issuance.

\section{Revocation and Time-Bound Access}

Access in Zero Trust CI/CD must be revocable—not just based on token expiration but based on runtime policy evaluation. Unlike traditional systems where access is granted once and assumed valid for its duration, intent-aware brokers must continuously re-evaluate whether access is still justified.

\subsection*{The Challenge with JWTs and Certificates}

\textbf{JWTs (such as SPIFFE JWT-SVIDs)} are designed for self-contained identity, but do not support revocation natively. Once issued, they remain valid until expiry. To mitigate this:
\begin{itemize}
  \item Brokers must issue JWTs with \textbf{short TTLs} (e.g., 5–15 minutes).
  \item Brokers must \textbf{refuse to reissue} new tokens when justification changes.
\end{itemize}

\textbf{Certificates (X.509-SVIDs)} can technically be revoked using CRLs or OCSP, but these mechanisms are operationally complex and not suitable for short-lived CI workloads. Instead, we favor revocation-by-denial: stop issuing new tokens if policy checks fail.

\vspace{0.8em}
\noindent\textbf{Intent-Aware Lease Pattern:}

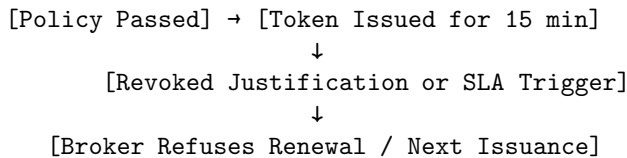
\begin{verbatim}
[Policy Passed] → [Token Issued for 15 min]
                      ↓
       [Revoked Justification or SLA Trigger]
                      ↓
   [Broker Refuses Renewal / Next Issuance]
\end{verbatim}

\subsection*{Broker Simulation in Python}

\begin{lstlisting}
def evaluate_access(request):
    if request["justification"] != "approved":
        return None
    if not is_within_window(request["timestamp"]):
        return None
    return issue_token(request["spiffe_id"])

def is_within_window(ts):
    return "02:00" <= ts <= "05:00"
\end{lstlisting}

\textbf{Key point:} revocation is enforced not by invalidating tokens mid-use, but by refusing further issuance when runtime signals change.

\subsection*{Credential Issuance Example: AWS STS with SPIFFE Identity}

A frequent implementation is a broker issuing short-lived AWS credentials using a SPIFFE JWT-SVID and the \texttt{AssumeRoleWithWebIdentity} API. This allows SPIFFE identities to interoperate with cloud-native identity systems in a secure, revocable way.

\textbf{Broker Flow:}

\begin{verbatim}
[CI Job] → [SPIRE issues JWT-SVID] 
            → [Broker evaluates policy: justification + time]
                → [Broker calls AWS STS AssumeRoleWithWebIdentity]
                    → [Job receives 15-min credentials]
\end{verbatim}

\textbf{Python Example:}

\begin{lstlisting}
import boto3

def assume_aws_role(jwt_token):
    sts = boto3.client('sts')
    creds = sts.assume_role_with_web_identity(
        RoleArn="arn:aws:iam::123456789012:role/CIProdDeployRole",
        RoleSessionName="ci-session",
        WebIdentityToken=jwt_token,
        DurationSeconds=900
    )
    return creds['Credentials']
\end{lstlisting}

\textbf{Security Characteristics:}
\begin{itemize}
  \item Token lifetime is tightly bounded (15 minutes).
  \item No static secrets or permanent role bindings.
  \item Future requests can be rejected based on live policy state.
\end{itemize}

\subsection*{Runtime Signals That Trigger Revocation}

\begin{itemize}
  \item SLA state changes from \texttt{normal} to \texttt{critical}
  \item Human override is expired or withdrawn
  \item Change ticket marked \texttt{rejected} or \texttt{rollback}
  \item Deployment window closes
\end{itemize}

By encoding these signals into broker logic or policy engines like OPA or Cedar, Zero Trust CI/CD environments maintain fine-grained control over workload access.

\subsection*{Summary}

Revocation in this architecture is implemented as \textbf{denial of further issuance}, rather than attempting to forcibly revoke a token already issued. Combined with short TTLs and tight policy checks, this results in a safer, more controllable authorization flow suited to ephemeral CI/CD workloads.

Next, we explore how human-in-the-loop approval systems can inject intent into these policy decisions via justification tokens and policy metadata.

\subsection*{Sample Justification Token Payload}

A justification token may be passed from an external system or stored in a credential broker cache. Here's an example of a signed approval token payload:

\begin{lstlisting}
{
  "token_id": "change-req-2025-112",
  "status": "approved",
  "approver": "release-manager@example.com",
  "issued_at": "2025-04-18T21:05:00Z",
  "expires": "2025-04-19T03:00:00Z",
  "reason": "Emergency patch to fix SLA breach",
  "source": "pagerduty"
}
\end{lstlisting}

Brokers pass this token (or a reference to it) into the policy engine as part of the evaluation context. It serves as a time-bound, human-issued justification for elevated access.

\section{Justification Tokens and Human-in-the-Loop Access}

While SPIFFE-based identity establishes \textit{who} is making the request, it does not capture \textit{why}. In high-assurance environments—particularly for production deployments, incident mitigation, or sensitive data access—authorization must include a layer of human intent and approval.

\subsection*{What is a Justification Token?}

A justification token represents external authorization metadata that a policy engine can inspect. It could be:
\begin{itemize}
  \item A signed approval object from a change management system
  \item A token issued by PagerDuty for an incident response
  \item A temporary override key generated via Slack, ServiceNow, or internal approval tools
\end{itemize}

These tokens are passed into the broker context and validated during policy evaluation.

\vspace{0.8em}
\noindent\textbf{Authorization Flow with Justification:}
\begin{verbatim}
[CI Job] 
  ↓
[SPIFFE Identity] + [Justification Token]
  ↓
[Broker → Policy Engine (OPA/Cedar)]
  ↓
[Access Granted or Denied]
\end{verbatim}

\subsection*{Cedar Example with Justification}

\begin{lstlisting}
permit(
  principal == workload::"spiffe://ci/org/critical-deploy",
  action == action::"deploy",
  resource == cluster::"prod-east"
)
when {
  context.approval == "change-req-2025-112" &&
  context.token.valid == true &&
  context.token.expires >= now()
};
\end{lstlisting}

\subsection*{OPA Rego Example}

\begin{lstlisting}
package authz

default allow = false

allow {
  input.spiffe_id == "spiffe://ci/org/deploy"
  input.approval.status == "approved"
  input.token.expiry > time.now
}
\end{lstlisting}

\subsection*{Broker Runtime Example (Python)}

\begin{lstlisting}
def check_approval(input_data):
    token = input_data.get("token", {})
    if token.get("status") != "approved":
        return False
    if token.get("expiry") < datetime.utcnow():
        return False
    return True
\end{lstlisting}

\subsection*{Human-in-the-Loop Use Cases}

\begin{itemize}
  \item \textbf{Production Release Approval}: CI jobs blocked until a release manager issues a signed token.
  \item \textbf{Incident Patching}: PagerDuty generates temporary override tokens during an outage window.
  \item \textbf{Maintenance Windows}: Dev teams submit maintenance plans that generate scoped tokens valid for a time period.
\end{itemize}

\subsection*{Design Considerations}

\begin{itemize}
  \item Tokens should be signed or stored in an auditable approval registry.
  \item Brokers must be able to query or verify tokens independently.
  \item The expiration or status of the token must influence real-time access decisions.
\end{itemize}

\subsection*{Summary}

Justification tokens inject human intent into machine-time authorization. They serve as run-time-bound approval attestations, enabling safe override, emergency access, and sensitive deployment gating. In intent-aware CI/CD systems, they are essential for bridging operational workflows with policy engines.

Next, we explore full-stack architectural patterns that tie together identity, broker logic, and runtime policy enforcement.

\section{Architectural Patterns}

Intent-aware authorization introduces a modular architecture where identity, policy, and access enforcement are fully decoupled. This separation of concerns supports composability, extensibility, and secure delegation in dynamic CI/CD environments.

\vspace{0.5em}
\noindent\textbf{Reference Flow:}
\begin{verbatim}
[SPIFFE Identity] 
       ↓
[Credential Broker] 
       ↓
[Policy Engine (OPA/Cedar)] 
       ↓
[Token Issuer / Cloud Access Gateway]
       ↓
[Target System (Cloud API, Secret Store)]
\end{verbatim}
\vspace{0.5em}

This layered design ensures that identity issuance (via SPIFFE), authorization logic (OPA/Cedar), and credential generation (STS, Vault, DB proxies) can evolve independently.

\subsection*{Push vs Pull Enforcement Models}

\textbf{Push Model:} Policy bundles are precompiled and pushed to the broker. This offers speed but limits runtime sensitivity.

\textbf{Pull Model:} Brokers evaluate fresh policies with contextual input per request. This model supports dynamic signals like SLA state, justification tokens, and time windows.

We observe that:
\begin{itemize}
  \item Push models suit high-throughput services (e.g., internal APIs)
  \item Pull models excel in CI/CD pipelines, where access is short-lived and context-heavy
\end{itemize}

\subsection*{Delegation Across Tenants and Teams}

In platform engineering environments, brokers are often shared across hundreds of teams. Identity scoping and policy boundaries become critical:

\begin{verbatim}
spiffe://org/team-alpha/deploy
spiffe://org/team-beta/build
\end{verbatim}

Per-tenant policies can be enforced by:
\begin{itemize}
  \item Path-based scoping (prefix match)
  \item SPIRE selectors (e.g., region, team, SLA class)
  \item Metadata tags passed to OPA/Cedar
\end{itemize}

This supports secure multi-tenancy without requiring isolated broker instances per team.

\subsection*{Trust Domains in Multi-Cloud Deployment}

Federated SPIFFE domains enable identity verification across environments:

\begin{verbatim}
[CI Job in GCP]
    ↓
[Broker in AWS]
    ↓
[Federated Trust Bundle]
    ↓
[OPA Policy + AWS STS Credential]
\end{verbatim}

Intent-aware architectures allow brokers to issue cloud-native credentials while respecting local policy and external trust anchors.

\section{Implementation Notes}

\subsection*{OPA: Context-Rich Policy Evaluation}

OPA policies are written in Rego and support JSON input. This makes it easy to pass SPIFFE IDs, deployment metadata, justification tokens, and time-based attributes.

OPA supports:
\begin{itemize}
  \item REST API queries
  \item Sidecar integration (e.g., with SPIRE Agent)
  \item Bundle distribution with versioned policy packs
\end{itemize}

OPA is especially effective in event-driven CI/CD brokers due to its small runtime footprint and clear audit logs.

\subsection*{Cedar: Hierarchical and Graph-Based Policies}

Cedar enables fine-grained, object-oriented access control using:
\begin{itemize}
  \item Entity relationships (e.g., job → service → cluster)
  \item Conditional clauses (e.g., SLA state, token validity)
  \item Multi-level delegation
\end{itemize}

It supports structured ABAC and RBAC hybrid models, making it suitable for environments with shared resources and compliance constraints.

\subsection*{SPIRE Integration and Enrichment}

SPIRE supports custom attestors and selector plugins to add:
\begin{itemize}
  \item Region or cloud provider metadata
  \item SLA tiers (e.g., bronze, silver, gold)
  \item Tags like \texttt{pipeline\_type=release}
\end{itemize}

These selectors become part of the SPIFFE ID issuance context and are passed to downstream policy systems.

\section{Related Work}

\subsection*{Standards and Frameworks}

\textbf{NIST SP 800-204} outlines strategies for microservices authentication and dynamic trust establishment. This paper extends those principles to CI/CD pipelines and non-human identity governance \cite{nist-sp800204}.

\textbf{WIMSE (IETF Working Group)} focuses on workload identity interoperability across domains. Our use of SPIFFE and trust bundles builds directly on these efforts.

\textbf{SPIFFE Federation and Runtime Brokers} were explored in our earlier work \cite{avirneni2025spiffe, avirneni2025broker}. This paper deepens the policy enforcement and intent modeling layers.

\subsection*{Industry Systems and Research}

\textbf{Aembit} and other Workload IAM solutions deliver similar runtime policy evaluation, but are often closed-source or cloud-specific. Our architecture uses portable, vendor-neutral building blocks.

\textbf{Google Zanzibar} and \textbf{Cedar} present scalable, policy graph engines. Cedar is especially relevant as it is OSS and tailored for structured authorization across cloud-native platforms.

\textbf{Credential Revocation} has long been a challenge in X.509 and JWT-based systems. Our use of control loop denial models and time-bound token leasing offers a lightweight, enforceable alternative.

\section{Discussion: Threat Model, Limitations, and Evaluation}

\subsection*{Threat Model}
We assume an adversary with access to compromised CI jobs, rogue users with valid SPIFFE identities, or unauthorized reuse of expired credentials. Brokers and policy engines are trusted execution components. The system assumes availability of external approval systems (e.g., PagerDuty, ServiceNow) and audit trails.

Potential attack vectors include:
\begin{itemize}
  \item Forged justification tokens if signing systems are misconfigured.
  \item Misconfigured policy rules leading to unintended access.
  \item Denial-of-service or latency issues in high-throughput CI/CD environments.
  \item Broker compromise, which could lead to policy bypass or unauthorized credential issuance.
\end{itemize}

\subsection*{Limitations}
\begin{itemize}
  \item Continuous policy evaluation may introduce latency in tight CI job windows.
  \item External systems must be highly available for justification checks.
  \item Cedar and OPA policy debugging may be nontrivial for large organizations.
\end{itemize}

\subsection*{Operational Usability}
We note that large organizations benefit from standardizing justification flows through structured tokens, while smaller orgs may prefer human approvals tied to GitHub PR metadata or Slack commands. Failure modes (e.g., approval system down) should default to deny, but configurable fallback logic may be used where appropriate.

\subsection*{Comparison Table: Authorization Models}
\begin{table}[H]
\centering
\caption{Comparison of Access Control Models}
\begin{tabular}{|l|c|c|c|}
\hline
\textbf{Feature} & \textbf{RBAC} & \textbf{ABAC} & \textbf{Intent-Aware} \\
\hline
Role/Tag-Based & Yes & Partial & No \\
Runtime Context & No & Limited & Yes \\
Human Approval Integration & No & No & Yes \\
Revocation Sensitivity & No & Limited & Yes \\
Justification Evaluation & No & No & Yes \\
Suitable for CI/CD & Limited & Medium & High \\
\hline
\end{tabular}
\end{table}

\section{Conclusion}

Intent-aware authorization is the natural evolution of Zero Trust CI/CD. It moves beyond "who are you?" into "why do you need access, now, under what conditions?" This shift is essential in environments where pipelines are short-lived, access is sensitive, and risk is dynamic.

By combining:
\begin{itemize}
  \item \textbf{SPIFFE-based identity} (verifiable, runtime-issued)
  \item \textbf{Credential brokers} (decoupling identity from access)
  \item \textbf{Policy engines} (OPA, Cedar) with context and justification
\end{itemize}

...we achieve secure, auditable, and revocable access at runtime.

\vspace{1em}
\noindent\textbf{Key Takeaways:}
\begin{itemize}
  \item Access should be granted not just by role or tag, but based on declared intent.
  \item Authorization must be modeled as a runtime control loop with feedback.
  \item Human approvals, SLA triggers, and incident context must influence policy.
  \item Brokers become enforcement meshes—contextual, continuous, and verifiable.
\end{itemize}

\vspace{1em}
This paper concludes a three-part series on Zero Trust CI/CD. While identity and access are foundational, we argue that \textbf{intent} is the missing dimension—where compliance, governance, and security meet operational velocity.

These patterns extend beyond CI/CD pipelines. They are equally relevant to any platform engineering environment that supports multi-tenant workloads, federated systems, or infrastructure-as-a-service platforms. Embedding intent into policy decisions enables scalable enforcement without central bottlenecks—key to building reliable and secure internal platforms at scale.

\end{document}